\documentclass[11pt,preprint]{aastex}
\usepackage{times}
\def\etal{{et al.}}

\def\eff{e\!\;\!f\!f}

\def\gtwod{g_{2D}}
\def\gtwodeff{g_{2D\!,e\!\;\!f\!f}}
\def\excap{\mbox{\scriptsize excised cap}}
\def\exmap{\mbox{\scriptsize excised map}}
\def\cap{\mbox{\scriptsize cap}}
\def\map{\mbox{\scriptsize map}}

\begin{document}

\title{Genus Topology of the Cosmic Microwave Background from WMAP}
\author{Wesley N. Colley\altaffilmark{1}, and
J. Richard Gott, III\altaffilmark{2}}

\altaffiltext{1}{Dept. of Astronomy, University of Virginia, P.O. Box
3818, Charlottesville, VA 22903}
\altaffiltext{2}{Dept. of Astrophysical Sciences, Princeton University,
Peyton Hall, Ivy Lane, Princeton, NJ 08544}

\begin{abstract}
We have independently measured the genus topology of the temperature
fluctuations in the cosmic microwave background seen by the Wilkinson
Microwave Anisotropy Probe (WMAP).  A genus analysis of the WMAP data
indicates consistency with Gaussian random-phase initial conditions,
as predicted by standard inflation.
\end{abstract}

\keywords{cosmology --- cosmic microwave background:  anisotropy}

\section{Introduction}

The greatly anticipated results from the WMAP project (Bennett \etal\ 2003a)
have redefined the state of the art in Cosmic Microwave Background (CMB)
science.  The WMAP project has provided full-sky coverage of the CMB at
unprecedented angular resolution.  The improvement over the only previous
full-sky dataset (COBE, Smoot \etal\ 1992) is of order 500 in terms of
resolution elements, and the sky coverage improvement over balloon and
ground-based experiments (e.g., de Bernardis \etal\ 2000 [Boomerang],
Stompor \etal\ 2001 [Maxima], Mason \etal\ 2002 [CBI], Kovac \etal\ 2002
[DASI]) is several hundred.  The experiment has resulted in dramatically
improved constraints on cosmological parameters, such as the matter
density, $\Omega_m$, Hubble Constant, $H_0$, and cosmological constant,
$\Omega_\Lambda$ (Spergel \etal\ 2003).

The tool of choice for assessing the cosmological parameters, is the power
spectrum, where one calculates the products of the spherical harmonic
coefficients $a_{\ell m}$ and their complex conjugates (for each $\ell$ and
$m$ in the $Y_{\ell m}$ expansion of the CMB sky, summed over $m$ values to
form $C_\ell$).  This $a_{\ell m}a_{\ell m}^*$ product, however, explicitly
removes phase information.  And while the power-spectrum is a very powerful
tool for assessing cosmological parameters such as $\Omega_m$ and
$\Omega_\Lambda$, it does (explicitly) remove any phase information
contained in the $a_{\ell m}$ coefficients themselves.  These phases,
however, contain critical information for characterizing the primordial
density fluctuations.  Namely, standard inflation (e.g., Guth 1981,
Albrecht \& Steinhardt 1982, Linde 1982, Linde 1983) predicts that the
temperature fluctuations in the CMB, at the resolution measured by WMAP,
will be characterized by spherical harmonic coefficients with Gaussian
distributed amplitudes and random phases.  The WMAP data provide our best
opportunity to date to test that hypothesis.

The genus topology method developed by Gott, Melott \& Dickinson (1986)
directly tests for the Gaussian random-phase nature of a density (or
temperature) distribution in 3 dimensions (Adler 1981; Gott, Melott \&
Dickinson 1986; Hamilton, Gott \& Weinberg 1986; Gott, Weinberg \& Melott
1987), or in 2 dimensions (Adler 1981; Melott \etal\ 1989). Coles (1988)
independently developed an equivalent statistic in 2 dimensions.  The 2
dimensional case has been studied for a variety of cosmological datasets:
on redshift slices (Park \etal\ 1992; Colley 1997), on sky maps (Gott \etal\
1992; Park, Gott, \& Choi 2001; Hoyle, Vogeley \& Gott 2002); and on the
CMB, in particular (Smoot \etal\ 1992; Kogut 1993; Kogut \etal\ 1996;
Colley, Gott \& Park 1996; Park, C-G. \etal\ 2001).

The WMAP team has recently measured the genus of the WMAP sky (Komatsu
\etal\ 2003), and demonstrated that the WMAP results are consistent with
the Gaussian random-phase hypothesis.  To do this, they carried out a large
number of simulations of the CMB, in which the spherical harmonic
coefficients were drawn from a Gaussian random-phase distribution.  They
then used their known beam profiles, to simulate the results in each
frequency, and applied the {\it Kp0} (Bennett \etal\ 2003b) mask, just as
one would with the real dataset.  For each of these simulations, they
computed the the genus (as defined by Melott \etal\ [1989]) and compared it
to the to that of the real dataset.  They found that the real dataset does
not depart significantly from the genus of the simulated Gaussian
random-phase datasets, very much in agreement with the results predicted
for WMAP by Park \etal\ (1998).  That group further explored Gaussianity by
testing other Minkowski functionals (Minkowski 1903) and the bispectrum
(related to the three-point correlation function), and found constraints on
Gaussianity similar to those provided by the genus study.

We seek, first, to confirm this result using our own methods, but also to
compare the genus measured from the data directly to the theoretical genus
curve for structures on a sphere.  Rather than simulating Gaussian
random-phase realizations of the CMB, we compare directly to the
theoretical prediction for the genus in two dimensions.  Furthermore, we
provide some details about the nuances of carrying out the genus
calculation on the HEALPix map projection and the stereographic map
projection.

\section{WMAP Observations}

The Wilkinson Microwave Anisotropy Probe (WMAP)\footnote{The WMAP homepage
is at http://map.gsfc.nasa.gov/} project team has released its one-year
dataset in several different skymaps measured at varying frequency and
angular resolution. Namely, the project has measured the anisotropy at
frequencies of 23, 33, 41, 61, and 94 GHz at angular resolutions of 0.82,
0.62, 0.49, 0.33, and 0.21 degrees (FWHM beamwidths), respectively. 

The WMAP team has released the data in a unique map projection called the
HEALPix projection (G\'orski \etal\ 2000).  We have chosen to re-present
the WMAP results in Fig.~\ref{fullstereo} for a few reasons.  First, we
will be studying the genus on stereographic projections (sections 6 and 7),
so it is useful to see the full sky map in that projection.  Second, we
have chosen a much different color scheme that is more relevant to genus
studies than the original color scheme used by the WMAP team.  The WMAP
color scheme is pretty, informative and impressive, but is not symmetric
with respect to hot and cold spots.  In our color scheme, the mean
temperature contour is white.  Higher temperatures than the mean are
represented as linearly redder with temperature.  Lower temperatures than
the mean are represented as linearly bluer with temperature.  Therefore,
the amount of blue ink per pixel on the page is linearly proportional to
the coldness of a cold spot, and the amount of red ink per pixel is
linearly proportional to the the hotness of a hot spot.  The scale runs
from $-200\mu\mbox{K}$ (solid blue) to $+200\mu\mbox{K}$ (solid red), the
same range as in the temperature maps produced by the WMAP team (Bennett
\etal\ 2003a). Since the two-dimensional genus of a Gaussian random-phase
field is all about the symmetry between hot spots and cold spots, we find
this color map highly instructive for our purposes.

\section{Genus Topology on a Sphere}

The properties of the genus are well-known in three dimensions (3D) (Gott
\etal\ 1986, Hamilton \etal\ 1986, Gott \etal\ 1987), but require some
explanation in the two-dimensional (2D) case (Melott \etal\ 1989),
particularly in the case of the sphere (Gott \etal\ 1990).

For 2D topology on a plane the 2D genus of a microwave background map is
defined as (Melott \etal\ 1989):
\begin{equation}
\gtwod = \mbox{Number of hot spots} - \mbox{Number of cold spots}
\end{equation}
For a Gaussian random field,
\begin{equation}
\gtwod \propto \nu\exp(-\nu^2/2),
\label{g2d}
\end{equation}
where $\nu$ is a parameter that measures the area fraction in the hot spots:
\begin{equation}
f = (2\pi)^{-1/2} \int^\infty_\nu{\exp(-x^2/2)dx}.
\label{areafrac}
\end{equation}

So for $\nu > 0$ ($f < 0.5$) there are more hot spots than cold spots,
while for $\nu < 0$ ($f > 0.5$) there are more cold spots than hot spots.
The genus is also equal to the integral of the curvature around the
temperature contour divided by $2\pi$.  If we were to drive a truck around
an isolated hot spot we would have to turn a total angle of $2\pi$ as we
complete an entire circuit around the hot spot. Driving a truck around an
isolated cold spot, we would turn a total angle of $2\pi$ with the opposite
sign, with a negative turn angle defined as one that is a turn to the left
when the hot region is on your right.  Thus, in a temperature field that is
divided into pixels we may define a pixel as hot if it is above the contour
threshold and cold if it is below the contour threshold.  Then the contour
line is actually composed of a series of line segments with turns occurring
at vertices in the pixel map.  There is a certain turn occurring at each
vertex (which, divided by $2\pi$, is equal to the contribution to the
genus).  Around each vertex there are four pixels, the contribution to the
total turn (and genus) at that vertex depends on how many of the four
surrounding pixels are hot and cold:

\begin{tabular}{llll}
0 hot, 4 cold& turn = 0, genus = 0 & the contour line does not intersect
the vertex\\
1 hot, 3 cold& turn = $90^\circ$, genus = $+1/4$& the contour line makes a
right angle turn\\
2 hot, 2 cold& turn = 0, genus = 0& no turn, or turns that add to zero on
average\\
3 hot, 1 cold& turn = $-90^\circ$, genus = $-1/4$& right angle turn around a
cold spot\\
4 hot, 0 cold& turn = 0, genus = 0& contour line does not intersect the vertex. \\
\end{tabular}

This is carried out by an identical algorithm to that in the CONTOUR2D
program as described in Melott \etal\ (1989).

Now we wish to define rigorously the 2D genus on a spherical surface. Here
we will follow our previous derivation (Gott \etal\ 1990).  The 2D genus is
defined to be equal to minus the 3D genus of the hot regions confined
within a thin spherical shell, measured in 3D.  Use lead paint to paint the
hot regions on the surface of a balloon---burst the balloon and you will
have solid curved lead shapes that will have a certain 3D genus---take the
minus of this number and that will be the 2D genus as we will define it.

Recall that we have defined the 3D genus as the number of (doughnut) holes
minus the number of isolated regions.  Our 3D genus is equal to the
integral of the Gaussian curvature over the contour surface divided by
$-4\pi$.  Thus, a sphere in 3D has a genus of $-1$, because it is one
isolated region.  The Gaussian curvature is $+(1/r^2)$ while the area of
the sphere is $4\pi r^2$, so the integral of the Gaussian curvature over
the sphere is $4\pi$, independent of the size of the sphere, and the genus
is $-1$.  A cube has a genus of $-1$ also, because it has the same topology
as a sphere.  In this case, the curvature of the surface entirely consists
of 8 delta functions at the eight vertices where three squares meet at a
point introducing a contribution to the Gaussian curvature integral of
$\pi/2$ at each vertex, equal to the conical deficit angle at the
vertex. (If we sandpapered off the corners and edges of the cube, we would
produce at each vertex an octant of a sphere containing a contribution to
the Gaussian curvature integral of 1/8th that of a sphere or $\pi/2$.  The
edges would be sanded off to quarter cylinders which have no Gaussian
curvature and the faces would have no Gaussian curvature, so all the
curvature is located at the vertices).  A doughnut has a genus of 0,
because it has one hole and consists of one isolated region.  A sphere with
two handles has a genus of $+1$ on our definition because it consists of
one isolated region and has two holes.

So using these definitions we can define the genus of a microwave
background map on the celestial sphere.  Suppose for example we have one
hot spot in the north polar region, and the rest of the sphere is cold.
Then the genus would be $+1$, because the hot spot cap is one isolated
region.  Now imagine the hot region covers the northern hemisphere while
the southern hemisphere is cold.  The genus would still be $+1$, because a
hemispherical bowl is one isolated region.  Suppose the hot region covers
all of the sphere except for a cold spot in the south polar region.  The
genus would still be $+1$, because this would look like a sugar bowl with
out any handles, which is also one isolated region.  The topology in each
case is identical since one can be deformed into the other, so the genus
should be the same in all three cases.  If the hot region ran around the
equatorial regions and there were cold spots at the north and south polar
regions, the 2D genus would be 0 because in 3D the equatorial band
(confined in a thin spherical shell) is a doughnut in 3D which has a 3D
genus of 0.  If the sphere had $N$ isolated hot spots (polka dots) on a
cold background, the 2D genus would be $N$.  If the sphere had $N$ isolated
cold spots on a hot background the genus would be $2-N$.  (Remember that 1
isolated hot spot on a cold background [a hot spot centered on the north
pole; the rest cold] is the same topologically as an isolated cold spot
centered on the south pole [the rest hot], because the latitude of the
contour line can be simply moved downward to transform one into the other.)
Suppose we have a hot spot covering the northern hemisphere with the
southern hemisphere cold.  This is a spherical cap of area $2\pi$
steradians.  Its 2D genus is $+1$ because in 3D it is one isolated region.
Considered as a spherical cap (part of a thin spherical shell) this
isolated region must have an integral of the Gaussian curvature on its
surface equal to $4\pi$ by the Gauss-Bonet theorem.  There are several
contributions to this: there is a contribution of $2\pi$ from the outside
(spherical) cap surface, and a contribution of $2\pi$ from the inside
(spherical) cap surface (a spherical shell has an inside and an outside
boundary both spherical), and a net contribution of 0 from the delta
function curvatures on the inside and outside edges at the equator marking
the boundaries of the inner and outer spherical caps---the contribution of
the outer edge is positive, while the contribution of the inner edge is
saddle-shaped and therefore negative.  The equator is a geodesic, so we
could drive a truck around the equator without turning; hence, the integral
of the turning around the contour boundary is zero.  Since we are circling
an isolated hot spot, why is the total turning not $2\pi$ as it is on the
plane?  Because we are circling a curved region, and if we parallel
transport (without turning) a vector around the boundary it will suffer a
deflection equal to the integral of the Gaussian curvature in the interior
of the boundary, which in this case is $2\pi$, the value we expect for a
closed curve in the plane.  The vector must return to its initial position
after circling, so the total of the turning integral and the parallel
transport deflection due to circling a curved region must add to give
$2\pi$.

Now consider a hot spot centered at the north pole and whose boundary
extends to latitude $30^\circ\mbox{N}$.  This is a spherical cap with area
$\pi$.  It is an isolated region so the 3D genus is $-1$, and the sum of
the integral of the Gaussian curvature over the spherical cap in 3D must be
$4\pi$.  The integral on the outside of the spherical cap is $\pi$, and on
the inside of the spherical cap is also $\pi$, and the boundary surface
connecting the inner and outer spherical cap shells is locally flat (part
of a cone), so the contribution from the inside and outside edges of the
boundary must be $2\pi$, which must be equal to twice the turning integral
for a truck driving on the spherical surface (since the total Gaussian
integral, $4\pi$, is twice the $2\pi$ value for circling an isolated
boundary [turning plus deviation from parallel transport]).  If we were to
parallel transport a vector around this boundary, we would find a
deflection of $\pi$ because that is the area in the spherical cap; to
return to its original orientation, a rotation of $2\pi$, the truck must
turn an additional angle of $\pi$.  Indeed, if you would like to drive
counterclockwise around the sphere at latitude $30^\circ\mbox{N}$, you will
have to keep your steering wheel turned to the left, since this not a
geodesic, and the total of that turning integral is $\pi$.  Now, locally it
is the turning integral that is being calculated from the formula for a
Gaussian random field.  So let us define the effective genus as
\begin{equation}
\gtwodeff  = \gtwod - 2f,
\end{equation}
where $f$ is the fraction of the area of the sphere in the hot spots.  In
our example, where the one hot spot was the entire northern hemisphere,
$\gtwod = 1$, $f = 1/2$ and $\gtwodeff = 0$.  The value of $\gtwodeff$
is equal to the turning integral of a truck driving around all the
individual components of the boundary contour (around all the hot spots)
divided by $2\pi$; in this case it is 0 because the truck driving around
the equator drives straight ahead without turning.  For the case where the
hot spot is centered at the north pole and extends to latitude
$30^\circ\mbox{N}, f = 1/4$, so the value of $\gtwodeff = 1/2$, because
the turning integral in this case is $\pi$.  Now for a Gaussian random
field on the sphere
\begin{equation}
\gtwodeff \propto \nu\exp(-\nu^2/2),
\end{equation}
because the Gaussian random field locally behaves on the sphere as on the
plane to produce this contribution to the turning integral.  Thus, in
comparing the WMAP data to the random-phase formula we will use $\gtwodeff$
defined rigorously as described above.

\section{Genus in the HEALPix Projection}

The WMAP data are plotted using an unusual map projection of the sphere,
called the HEALPix projection (G\'orski \etal\ 2000).  The sphere is
effectively projected onto a rhombic dodecahedron. (This semi-regular
polyhedron is the cell for the face centered cubic crystal lattice, the
polyhedron representing the boundary of the set of points closer to a
particular atomic nucleus than to any other.)  It has 12 faces, four
diamonds that meet at the top like the four sides of a pyramid, four
diamonds that circle the equator, and four that meet at the bottom like the
sides of an inverted pyramid.  If we were to set this rhombic dodecahedron
up aligned with the cardinal directions we could label the 12 faces as
follows: TN (top north), TE (top east), TS (top south), TW (top west), NE
(northeast, on the equator), SE (southeast, on the equator), SW (southwest,
on the equator), NW (northwest, on the equator), BN (bottom north), BE
(bottom east), BS (bottom south), BW (bottom west).  There are 6 vertices
where 4 diamonds meet at a point (i.e., TN, TE, TS, TW all meet at the top,
and similarly TN, NE, NW, BN meet at a vertex).  There are 8 vertices where
3 diamonds meet at a point (e.g., TN, TE, NE meet at a vertex).  Each
diamond may be subdivided into diamond shaped pixels.  This is done by
factors of 4, subdividing each diamond into 4 diamonds, and repeating by
similarly subdividing each diamond.  (In a similar way we may produce a
checkerboard of 64 square pixels by dividing a square first into 4 quadrant
squares, dividing each of these into 4 sub-squares, and each of these
sub-squares into 4 checkerboard squares). In the WMAP case these are all
diamonds (rhombuses) instead of squares, but the topology is the same.  The
subdivisions are made so that all the pixels are equal area, and diagonal
rows of pixels touching point to point have centers that lie on circles of
latitude.  With $N$ successive subdivisions of the diamonds into 4
sub-diamonds, we have a total of $12\times 4^N$ pixels of equal area
covering the sphere.  For example, WMAP uses $N = 9$, where there are
3,145,728 diamond shaped pixels covering the sphere, each with dimensions
of approximately $0.11^\circ \times 0.11^\circ$, which is adequate
resolution to show the WMAP data which has an angular resolution of
approximately $0.3^\circ$.  To study the topology properly, one needs
pixels that are at least 2.5 times smaller than the smoothing length.

Each of these diamond pixels on the sphere has angles which would be
complicated to calculate, so a version of CONTOUR2D for the sphere would
seem to be complex and require many calculations to calculate the turning
angles.  However, we may simplify this greatly by using the topological
invariance of $\gtwod$ to projection (first from the sphere to the rhombic
dodecahedron, then from the rhombic dodecahedron to a cube).  This cube
has six faces, which we will label T (top), N (north), E (east), S (south),
W (west), B (bottom).  Each diamond of the rhombic dodecahedron is
oriented with its short diagonal along an edge of the cube, and each of
the 12 rhombuses is mapped into two triangles that appear on adjacent faces
of the cube.  For example the rhombus TN is divided onto two triangles, one
of which is mapped onto the T map on the cube and the other is mapped onto
the N face of the cube.  In fact, the letter designation of each rhombus
tells us which of the two cubic faces it is mapped onto.  For example, the
SW rhombus is divided into two triangles which are mapped onto the S and W
faces of the cube.  The T face of the cube thus is a square which is
divided into 4 right triangles whose hypotenuses form the sides of the
square, and which meet in the center; these 4 triangles represent the upper
halves of the four rhombuses TN, TE, TS, TW which meet at the top.

In such a projection we paint the hot regions with lead paint on the surface
of the cube, and the 3D genus of these lead shapes is the same as if they
were painted onto the sphere, because they have only been distorted in
shape, not changed in topology.  The total integrated Gaussian curvature
over the cube, which totals $4\pi$ is concentrated entirely in 8 delta
functions at the 8 vertices, where there is an angle deficit of $\pi/2$ at
each, as 3 squares instead of four meet at each of these 8 vertices.  Each
of the 12 rhombuses of the original dodecahedron is mapped onto two right
triangles meeting at their hypotenuses along an edge of the cube; these can
be flattened out to make a square, with square pixels.  So we have
equivalently, 12 square maps of $4^N$ square pixels each.  Thus, we can
consider each rhombus as a square map with square pixels and we can
calculate $\gtwod$ for each using CONTOUR2D.  Vertices that are within the
rhombus are included.  Vertices that are along an edge with the next
rhombus are shared according to a prescription where the vertices along the
top and right edges of the rhombus are assigned to it when laid out
properly, and the bottom and left edges are assigned to adjacent rhombuses,
since two rhombuses meet along each edge.  That accounts for all pixel
vertices except for those at the corners of the rhombuses.  Each rhombus
has 2 corners where 4 rhombuses meet (such as TN, TE, TS, TW meet at the
top) and 2 corners where 3 rhombuses meet at a vertex (such as TN, TE, NE)
which occurs at a vertex of the cube where there is a $\pi/2$ angle deficit
in our cubic map and where 3 pixels only meet at a point.  The corners
where 4 rhombuses meet at a point are like ordinary vertices in the plane
where 4 square pixels meet at a point, so this is handled with CONTOUR2D.
The value of the contribution to the genus from that vertex must then be
divided and shared equally among each of the rhombuses that meet there.
The prescription for vertices where 3 pixels meet at a point (at the corner
of the cube in our projection) is as follows:

\begin{tabular}{ll}
0 hot 3 cold & genus = 0  the contour does not intersect the vertex \\
1 hot 2 cold & genus = 1/4  added to hot rhombus, 0 added to others \\
2 hot 1 cold & genus = 0 \\
3 hot 0 cold & genus = 1/12 added to each of the three rhombuses.\\
\end{tabular}

If we were to smooth the edges of the cube by sanding, the edges would turn
into quarter cylinders (with zero Gaussian curvature) and the 8 vertices
(where three squares meet at a point) turn into tiny octants of a sphere
each with an integrated Gaussian curvature of $\pi/2$.  The faces of the
cube are flat so all the curvature is located in the 8 vertices.  The edges
of the three rhombuses meet at $120^\circ$ angles in the spherical octant.
So in the case 1 hot and 2 cold there is a turn of $60^\circ$ there in the
contour plus a curvature of $30^\circ$ (equal to 1/3 of the integrated
curvature over the octant of the sphere) so the genus contribution (turn
plus parallel transport deviation = $90^\circ$) should correspond to $+1/4$
(or $\pi/2$ divided by $2\pi$).  This is then added to the hot rhombus as
this is the hot spot we are circling.  In the case 2 hot, 1 cold, there is
a $-60^\circ$ turn, plus two regions of $30^\circ$ curvature contributing
to the parallel transport deviation, making a total contribution of 0 to
the genus.  In the case 3 hot, 0 cold, the vertex lies entirely within the
hot region but it adds $\pi/2$ to the curvature integral within that region
and this must be shared equally with all three rhombuses to add $+1/12$ to
the genus in each.  The total genus for the sphere $\gtwod$ is calculated
by adding the genus from all 12 rhombuses.  From $\gtwod$ we can calculate
$\gtwodeff$ by subtracting $2f$.  We can calculate $f$ as simply the
fraction of the diamond shaped pixels that are hot, since these are all
equal area.  Thus, from the WMAP data we can calculate the 2D topology and
compare it with that expected from a Gaussian random-phase distribution.

\section{The Genus of the CMB, Measured by WMAP}

The most direct data product from the WMAP team is the ``internal linear
combination'' (ILC) map of the CMB, which is given in the HEALPix format.
This ILC map uses the optimum linear combination of the skymaps at the
different frequencies to remove the Galaxy and some other foregrounds
(Bennett \etal\ 2003b).  This is the primary map distributed as the best
rendering of the CMB and has a resolution (beam width) of $0.3^\circ \times
0.3^\circ$.  Though this map presents a few problems for directly computing
the genus (which we will address shortly), it is worth checking the genus
on this map directly.

The HEALPix projection provides a natural division of the dataset (sky)
into 12 independent regions.  We therefore measured the genus in each of
these regions and computed the total for the whole sky, which we plot in
Fig.~\ref{hpg_2}.  In this figure $\nu$ is computed in terms of the
pixel-wise temperature mean and standard deviation measured over the whole
sky (method 1).  This method would be equivalent to the area fraction
method described by equation (\ref{areafrac}) if the pixel histogram of
temperature were strictly Gaussian.  Fig.~\ref{hpg_1} is identical, except
that $\nu$ is computed by area fraction in each individual HEALPix rhombus
(method 2); see equation (\ref{areafrac}).  Using the area fraction method
measures more directly the random-phase nature of the distribution, since
it separates that from any departures from a Gaussian histogram of the
temperature (which could be measured directly).  In method 1, we use $\nu$
defined in terms of the standard deviation in temperature over the whole
sky; in method 2, we define $\nu$ by area fraction in each rhombus, thus
treating each of the 12 rhombuses completely independently.  The two
methods provide a contrast of data treatment, and yet, as we shall see,
give essentially identical results.

The inner errorbars in Figs.~\ref{hpg_2} and \ref{hpg_1} are the standard
deviation of the mean in the genus for the whole sky, estimated from those
the 12 independent regions of the sky (adding the values of $\gtwodeff$
obtained from each).  The standard one-sigma errorbars are shown as the
inner errorbars.  The outer errorbars show the $95.4\%$ (Gaussian
two-sigma) confidence interval for a Student's-$t$ variable with 12 degrees
of freedom, which we will explain shortly.

The solid curves in Figs.~\ref{hpg_2} and \ref{hpg_1} give the Gaussian
random-phase genus curve, according to equation (\ref{g2d}), with the best
fit amplitude applied.  The amplitudes, $A$, [$\gtwodeff =
A\nu\exp(-\nu^2/2)$], in Figs.~\ref{hpg_2} and \ref{hpg_1} are $A = 3432$ and
$A = 3657$, respectively.  These are by far the highest in any genus
measurements of cosmological structure to date.  As expected, the first
value of $A$ is lower than the second, because the power at large scales
(quadrupole and octopole) lifts some regions (rhombuses) to higher
temperature, congealing some hot spots therein, but lowers other regions
(rhombuses), congealing some cold spots therein.  Method 2 sets a median in
each rhombus separately and thus discovers more structures.

The Student's-$t$ formulation is necessary since we are estimating the
standard error with the distribution itself.  Namely, if the true genus for
the whole sky at each point on the curve were $\mu$ and the true standard
deviation in the mean expected for that number of structures were $\sigma$
(we expect the distribution to be Gaussian for Gaussian random-phase fields
[Gott \etal\ 1990]), then our estimator for $\mu$ is the arithmetic average
estimated from the 12 rhombuses (the values from each rhombus are
multiplied by twelve, then they are averaged), $\bar{x}$, and our estimator
for $\sigma$ is $s/\sqrt{n-1}$, where $s$ is the root-mean-square
difference from the mean (as usual).  We expect that $\mu$ is given exactly
by equation (\ref{g2d}) and that our best fit for the amplitude has
negligible error compared to any one point on the curve.  The expression
$(\bar{x}-\mu)/(s/\sqrt{n-1})$ is defined to be a Student's-$t$ variate
(Lupton 1993).

On average, we would expect only one of the 21 points in each of the genus
curves in Figs.~\ref{hpg_2} and \ref{hpg_1} to miss the curve outside of
the 95.4\% confidence errorbar.  In fact, not even one misses.  From the
properties of the Student's-$t$ distribution with 12 d.o.f., one would
expect to find that 33.7\% ($\sim 7$) of the 21 points miss the curve by
more than the inner errorbars.  In both cases, the agreement is again
better than expected, with only three misses in each figure.  This
better-than-expected agreement can easily arise by chance from correlations
between the genus measurements at neighboring values of $\nu$, as shown by
Colley (1997) (these correlations mean that the 21 data points are not
completely independent---neighboring genus values measure many of the same
structures).  Nonetheless, the genus of temperature fluctuations in the
CMB, as observed by WMAP, is certainly consistent with the Gaussian
random-phase hypothesis.

\section{Genus on a Stereographic Map Projection}

In the ILC HEALPix data release, there are a couple issues that
must be addressed.  The first issue is foreground emission.  Bennett \etal\
(2003b) discuss in quite a bit of detail the foregrounds in the
WMAP results.  They provide a very useful sky mask that should be applied
to avoid foreground contaminants, the principal one of which is the Galaxy.
In a second study, we have therefore used their {\it Kp0} (most
conservative) sky mask, which also removes some 200 point sources.
Furthermore, we have restricted ourselves to Galactic latitudes $|b| >
18^\circ$.  Since the mask must remove much of the area near the Galactic
Equator, the HEALPix projection, with its 4 equatorial diamonds, ceases to
be ideal.  Instead, we re-project the map, using the stereographic
projection for both the Northern and Southern Galactic Hemispheres (as in
Fig.~\ref{fullstereo}, in which the $|b| > 18^\circ$ cut occurs at 72.7\% of
the total radius in each hemisphere).

%Because galaxy contamination may still be worrisome, we may also want to
%look separately at north galactic and south galactic caps denoted by $b >
%18^\circ$ and $b < -18^\circ$ respectively.  This way the galaxy can be
%eliminated.  We can consider the co-added map as before or just look at the
%61GHz band separately, for example, which seems fairly free of
%contamination and is the primary contributor to the co-added map.

We can calculate $\gtwod$ from the north and south galactic caps by using
stereographic map projections of the caps. Stereographic map projections of
a hemisphere are conformal and preserve angles.  A stereographic projection
of the north galactic cap is a circle, in which azimuthal angle in the map
is the galactic longitude in the cap.  Radius in the map, $r =
2\tan[(90^\circ - b)/2]$ (See Fig.~\ref{fullstereo}).  The outside boundary
is a circle of radius $r_b = 2\tan[(90^\circ - 18^\circ)/2]$.  The contours
on the sphere can be approximated as spherical polygons---geodesic segments
joined at vertices where turning on the sphere occurs.  The stereographic
projection preserves angles, so the turning that occurs at vertices is
mapped properly.  But, geodesics (great circles) are mapped into arcs of
circles rather than straight lines, which turn on the map but not on the
sphere so the total genus $\gtwod$ on the sphere is not automatically
calculable from the total turning that comes from circling the density
contour on the map.

So instead of calculating $\gtwod$ for the sphere from the stereographic map
directly, which would be complicated, let us instead calculate the relation
between $\gtwodeff$ on a spherical cap, and $g_{2D\!,\mbox{\scriptsize
map}}$ measured by the CONTOUR2D program on the flat stereographic map of
the cap.  [Of course once we have $\gtwodeff$ we can calculate the
contribution to $\gtwod$ due to the structures in the north galactic cap by
adding $2f\cdot \mbox{Area}_{\cap}/(4\pi)$].  But it is
really $\gtwodeff$ we are interested in anyway because it is what will be
compared to the theoretical formula.  Thus, we can skip a step by going
directly after $\gtwodeff$.

To begin, let us excise the spherical cap from the sphere by cutting it off
and separating it from the rest of the sphere.  This will add boundary
terms.  Each hot spot (think curved lead sheet) that intersects the
boundary of the cap will be closed off by a curve along the boundary that
will make it an isolated hot spot.  Think about these hot spots in 3D when
the cap is excised from the rest of the sphere: any hot spot that hits the
boundary will become isolated, floating in space, because beyond the
boundary there is nothing.  This spherical isolated cap will have a value
of $g_{2D\!,\excap} = g_{2D\!,\exmap}$; the total genus of the isolated flat
stereographic map sitting isolated in the plane---think of flat lead sheets
lying in a plane; because one can be continuously deformed into the other.
We can calculate the terms contributing to $g_{2D\!,\excap}$:
\begin{equation}
\begin{array}{rl}
g_{2D\!,\excap} =
&  (2\pi)^{-1}\sum{(\mbox{turns on spherical cap})}\\
&+ f\cdot\mbox{Area}_{\cap}/(2\pi) \\
&+ (2\pi)^{-1}\sum{(\mbox{turns at cap boundary})}  \\
&+ f_b[1 - \mbox{Area}_{\cap}/(2\pi)].
\end{array}
\end{equation}
The first term, $(2\pi)^{-1}\sum{(\mbox{turns on spherical cap})}$, is
equal to $g_{2D\!,\eff,\cap}$ for the interior of the spherical cap.  It is the
total turns made driving on the sphere around all the contours within the
spherical cap, and excludes any boundary terms produced by excising the
spherical cap.  Now $g_{2D\!,\eff,\cap} \propto \nu\exp(-\nu^2/2)$, so it
is what we want to calculate to compare with the theoretical genus formula.
The term $f\cdot\mbox{Area}_{\cap}/(2\pi)$, where $\mbox{Area}_{\cap}$ is
measured in steradians, is the integrated Gaussian curvature within the hot
spots in the cap divided by $2\pi$, which contributes to the genus by the
3D argument presented above.  The term $(2\pi)^{-1}\sum{(\mbox{turns at cap
boundary})}$ is the sum of the turns that are made when you are driving
along a contour and you encounter the cap boundary and then must turn a
sharp corner there to continue driving along the cap boundary to complete
circling that hot spot as an isolated region.  In general, if along the
boundary there are $N$ hot segments, there will be $2N$ such turns, as we
enter and then exit each of these N hot segments along the boundary.  Each
of these $2N$ turns will be some angle between 0 and $2\pi$.  The final
term, $f_b[1 - \mbox{Area}_{\cap}/(2\pi)]$, represents the total turning on
the sphere that occurs driving along the $N$ hot segments of the boundary
of the cap.  If the entire boundary were hot then driving around it would
have a total turn of $2\pi - \mbox{Area}_{\cap}$.  (For example, if this
were a hemispherical cap the area of the cap would be $2\pi$ and the total
turn circling the boundary would be zero, because the boundary is the
equator and it is a geodesic so no turning is required to circle the
boundary).  As the area of the cap goes to zero, the curvature within the
cap becomes unimportant to parallel transport and the total turn required
to circle the boundary approaches $2\pi$.  In general, the angle deflection
suffered by parallel transport around the closed boundary curve is equal to
the integral of the Gaussian curvature inside, which in this case is equal
to the area of the cap.  When we have completed the circuit, we have
returned to where we have started so we have suffered a total rotation of
$2\pi$, so the turning we do on the sphere circling the boundary plus the
area of the cap is $2\pi$.  Thus, turning = $2\pi -
\mbox{Area}_{\mbox{\scriptsize cap}}$.  The number $f_b$ is the fraction of
the boundary that is hot, and so it is this fraction of the boundary that
we drive around.  Dividing this by $2\pi$, gives the contribution to the
genus derived by driving along the hot segments of the boundary: $f_b[1 -
\mbox{Area}_{\cap}/(2\pi)]$.  Using the Copernican Principle, one would not
expect the boundary to be special, so on average the fraction of the cap
boundary that was hot should be equal to the fraction of the sphere that is
in hot spots.  On average then, we expect $f_b = f$. Substituting we find:
\begin{equation}
g_{2D\!,{\excap}} = 
 		g_{2D\!,\eff,\cap} +
 		{{f\cdot \mbox{Area}_{\cap}}\over{2\pi}} +
 		{1\over{2\pi}}\sum{(\mbox{turns at cap boundary})} +
		f \left(1 - {{\mbox{Area}_{\cap}}\over{2\pi}}\right).
\end{equation}
\begin{equation}
g_{2D\!,\excap} = g_{2D\!,\eff,\cap} +
f + {1\over{2\pi}}\sum{(\mbox{turns at cap boundary})}.
\end{equation}
Now let us measure the genus on the excised stereographic map
($g_{2D\!,\exmap}$), recalling that the excised
stereographic map is topologically equivalent to the excised cap.
\begin{equation}
g_{2D\!,\exmap} = {1\over{2\pi}}\sum{(\mbox{turns on
stereographic map})} +
{1\over{2\pi}}\sum{(\mbox{turns at map boundary})} + f_b.
\end{equation}
The first term, $(2\pi)^{-1}\sum{(\mbox{turns on stereographic map})} =
g_{2D\!,\map}$, is the quantity measured by CONTOUR2D on the flat map.
CONTOUR2D examines only the Cartesian vertices within the stereographic map
region and does not include any boundary effects.  There is no
$f\cdot\mbox{Area}$ term because the map is flat and there is zero
curvature inside the hot spots.  The next term,
$(2\pi)^{-1}\sum{(\mbox{turns at map boundary})}$, adds the turns that are
taken when the contour hits the outer circular boundary of the map.  If we
use a stereographic projection, which is conformal and preserves angles,
each and every turn at the boundary in the map will be equal that
encountered on the sphere, so:
\begin{equation}
{1\over{2\pi}}\sum{(\mbox{turns at map boundary})}
= {1\over{2\pi}}\sum{(\mbox{turns at cap boundary})}.
\end{equation}
The final term, $f_b$, is the total turn taken on the $N$ hot segments of
the boundary of the map divided by $2\pi$.  If we circled the entire
boundary we would have to turn a total of $2\pi$, because the map is flat
and Euclidean geometry applies.  The $N$ hot segments cover a fraction
$f_b$ of the boundary (galactic longitude is mapped onto azimuth in the
stereographic map, so the fraction of the map boundary occupied by the $N$
hot segments is exactly the same as on the sphere).  Thus, the total turn
is $2\pi f_b$, and dividing this by $2\pi$ gives $f_b$.  As we have
remarked above, if the boundary is not special we expect $f_b = f$.  Since
the map is excised and sits alone on the plane, these outer boundary
segments must be included to create isolated hot spots in the plane out of
the hot regions that hit the boundary.  Equating $g_{2D\!,\mbox{\scriptsize
excised cap}}$ and $g_{2D\!,\exmap}$ (because the cap can be deformed into
the map), and substituting from the above equations we find:
\begin{equation}
g_{2D\!,\excap} = g_{2D\!,\exmap}.
\end{equation}
\begin{equation}
g_{2D\!,\eff,\cap} + f + {1\over{2\pi}}\sum{(\mbox{turns at cap boundary})} =
g_{2D\!,\map} + {1\over{2\pi}}\sum{(\mbox{turns at map boundary})} + f.
\end{equation}
\begin{equation}
g_{2D\!,\eff,\cap} = g_{2D\!,\map}.
\end{equation}
So ``what you see is what you get.''  If we measure the 2D topology in the
interior of the stereographic map using CONTOUR2D (which ignores the
boundary) our result will equal, on average, $g_{2D\!,\eff,\cap}$ which is
what we were looking for.  This works for any radius spherical cap, since
terms proportional to the area of the cap cancel out.  And since the map is
conformal, the turns at the boundary are equal in both cases and cancel out
as well.  Finally, since we expect the boundary to be typical, its coverage
with hot spots should be equal to the fraction $f$ of the sphere covered by
hot spots, and this causes other terms to cancel as well, leaving the above
simple result.

\section{Genus Topology on Stereographic WMAP Results}

We may now compute the 2-D genus on a stereographic re-projection of the
WMAP data, in which we address the issues of foreground contamination and
noise variability arising from the scanning strategy of the WMAP
spacecraft.

First, to alleviate contamination from the Galaxy, we completely exclude
the equator region up to $|b| < 18^\circ$.  The {\it Kp0} mask does include
removal of local regions for point source avoidance, and some additional
Galaxy masking above the $|b| > 18^\circ$ cut, particularly near the Galactic
Center, the Gum Nebula and Orion Complex.  Any vertex in the stereographic
map projection that falls within the mask is simply ignored by CONTOUR2D.

We proceed by dividing the stereographic projection into octants which
correspond to octants on the sphere.  These octants form 8 independent
regions in which to measure the genus.  A slight adjustment for the actual
solid angle in each octant is made, because the {\it Kp0} mask changes the
solid angle of each octant very slightly (a few percent variation).  We
normalize the genus to the mean solid angle of the masked octants, and
proceed treating each masked octant as one of 8 independent measurements of
the genus per average masked octant.

In carrying out the re-projection to the stereographic map, we have applied
a slight smoothing to a total smoothing FWHM of $0.35^\circ$.
Figs.~\ref{g.35_2} and \ref{g.35_1} show the genus for the masked
stereographic projection, computed by methods 1 and 2 (same methods as in
the HEALPix genus computation).  The errorbars are computed now for 8
octants instead of 12 diamonds, which changes somewhat the relative size of
the 95.4\% errorbars and the direct one-sigma errorbars.  The best fit
theoretical genus curve (equation \ref{g2d}) is plotted in each figure,
with respective amplitudes of $A = 2057$ and $A = 2186$ (as with the HEALPix
genus curves, the first value is slightly less than the second, as
expected).  The amplitudes have come down somewhat from the HEALPix maps
due to the exclusion of the equatorial band (so the total genus covers only
a fraction of the whole sphere), the additional masked pixels in the {\it
Kp0} prescription, and the additional smoothing applied in the
stereographic projection.

In each of the figures, the agreement with the theoretical genus curve is
striking.  Quantitatively, we would expect on average that one of the 21
points sampling the curve would miss the theoretical fit outside is 95.4\%
errorbar.  With method 1 (Fig.~\ref{g.35_2}), no points miss the
theoretical curve outside their 95.4\% errorbar, while with method 2
(Fig.~\ref{g.35_1}), two points miss by a small margin.  The one-sigma
(inner errorbars) for the 8 degrees of freedom of the octants reside at the
65.3\% confidence interval, so on average, one would expect approximately 7
of the 21 points on each of Fig.~\ref{g.35_2} and Fig.~\ref{g.35_1} to miss
the theoretical curve outside their one-sigma errorbars.  Using method 1,
we find that 3 points miss at this margin, and 9 miss using method 2.
Though the method 2 results show slightly more than the average number of
misses expected at both the one-sigma and 95.4\% level, the excess is
certainly within the uncertainty arising from small number statistics in
each case.

The second contaminant in the ILC HEALPix map is that the noise behavior is
very complicated, and differs substantially pixel to pixel.  First, the
beams at each frequency in the combination have different widths.  Second,
the scan method of the WMAP satellite yields many more samples at the
ecliptic poles than at the ecliptic equator (see Bennett~\etal\ 2003a,
Fig. 3).  To alleviate some of these problems, we have downloaded from the
WMAP website maps in the same 5 frequencies as used in the ILC map, but
ones that have been pre-smoothed to the same effective beam-width
($0.82^\circ$).  These are then added with the same weights as used in
producing the ILC map.  At cost of number of structures, these maps
significantly enhance the signal-to-noise of remaining structures, thus
reducing the significance of the noise differences arising from the
scanning strategy.  This is, therefore our most conservative map.

In projecting these maps into stereographic maps, we have smoothed to
$0.9^\circ$, slightly more than the $0.82^\circ$ smoothing of the
pre-smoothed maps.  Figs.~\ref{g.9_2} and \ref{g.9_1} show the genus
measured in these stereographic projections.  The best-fit amplitudes have
dropped to $A = 772$ for method 1 and $A = 831$ for method 2, due to the
smoothing.

From these final stereographic projections, the genus measured at the 21
points should also be expected to miss the theoretical curve outside of the
95.4\% errorbars once on each figure.  What we find is that with method 1
(Fig.~\ref{g.9_2}), no points miss the curve outside the 95.4\% errorbar,
while with method 2, (Fig.~\ref{g.9_1}), one point misses by a small
amount.  The points miss the curve outside their one-sigma errorbars 2
times for method 1, and 4 times for method 2, again a bit better than
expected.  The final figure, Fig.~\ref{g.9_1}, represents our most
conservative estimate of the true genus available in the WMAP distribution,
and shows excellent agreement with the Gaussian random-phase hypothesis.

\section{Discussion}

As we have shown, topologically, within the errors, the CMB measured by
WMAP appears to be consistent with Gaussian random-phase.  Indeed, the
overall visual impression is strikingly random-phase.  The stereographic
projection is conformal, so shapes are preserved locally, though the scale
grows as one approaches the the outer boundary (by a factor of two from the
pole to the equator, as seen in Fig.~\ref{fullstereo}).  We may ask if
there are any features visible in the map at all that look non-random-phase
in any way.

Our choice of map projection has a miscellaneous advantage over the
Molleweide projection used by the WMAP team, which is that it shows the
Galactic polar regions very well.  In the south polar region, there is one
possibly non-random feature, a fairly narrow red (hot) feature stretching
from approximately the 8:30 position toward the 2:30 position curving like
a bow below the South Galactic Pole.  This feature is more visually
apparent when viewed in landscape orientation.  It has the appearance of a
scanning error approximately along a great circle, but is not coincident
with any particular locus in the WMAP scanning geometry.

We have considered several possibilities to explain such a feature.  First,
there is the simple phenomenon of the ``canals'' on Mars, the linear
features reported in naked-eye observations of the red planet in the late
$19^{\mbox{\scriptsize th}}$ century.  We now know that there are no such
canals, and that our eyes' highly evolved ability to detect linear features
or edges was simply creating an illusion.  Most likely the same thing is
going on here, particularly since the feature is more apparent when the
figure is viewed in landscape rather than in portrait orientation.  In
fact, if one had a true Gaussian random-phase distribution and there were
no features that looked unusual to the eye, that would be unusual.  So this
feature may be merely coincidental.

What are the other possibilities?  It is not coincident with the Magellanic
Stream measured seen in 21cm (Brooks 2000).  Furthermore, we would expect
the stream to have similar microwave ``color'' to the gas in the Galaxy,
and hence be subtracted about as well as the Galaxy is in the ILC map.  To
test this, we constructed a three color map, using the K band as red, Q as
green and W as blue, and found that while the foregrounds were distinctly
pink to orange to yellow, the CMB was quite grey, and so was the feature.

Another possibility that comes to mind is cosmic strings.  Cosmic strings,
however, do not amplify temperature along a line.  Instead, a moving cosmic
string produces a cliff, hotter on one side (the trailing side) than the
other side (the leading side).  Thus a hot stripe would require two
parallel cosmic strings, having just passed each other (an interesting
theoretical possibility [Gott 1991], but not a likely one, since we are
unlikely to find the two most prominent cosmic strings within the horizon
in such an alignment by chance).  Also, there is no obvious evidence for
cosmic string cliffs anywhere else in the WMAP data, which we have tested
for by making gradient maps as suggested by Gott \etal\ (1990).

The feature could also be due to the integrated Sachs-Wolfe effect from a
great wall seen edge-on, but in looking at several galaxy catalogs and
X-ray maps, we could detect no obvious feature with appropriate geometry.

This list of possibilities is by no means exhaustive, and, in the end, it
may be most likely that we are simply falling prey to ``Martian canals'' in
an overall Gaussian random-phase map.  Since there are countless
possibilities for what one may see and deem non-random (circles, squares,
triangles, even letters or words), the {\it a posteriori} statistical
significance of any such weak feature is difficult to estimate.

The overall consistency with of the genus topology in the WMAP data with
that predicted {\it a priori} from a Gaussian random-phase distribution is
quite spectacular.  WMAP's unprecedented combination of angular resolution
and all-sky coverage of the Cosmic Microwave Background has presented by
far the best confirmation to date of the standard inflationary prediction
of Gaussian random-phase initial conditions.

In standard inflation, the structures seen in the CMB are fossil remains of
random quantum fluctuations occurring just $10^{-33}$ seconds after the Big
Bang.  In galaxy clustering surveys, these fossils have become somewhat
distorted due to non-linear effects and biased galaxy formation.  In the
CMB, however, we are seeing these fluctuations while they are still in the
linear regime, making the WMAP measurements of the CMB a very attractive
dataset for studying these primordial perturbations.

With the hundreds of structures observed even at the $0.9^\circ$ scale, the
WMAP results have provided the largest and best dataset for studying the
primordial structures.  The genus topology test, in particular, tests the
dataset for the standard inflationary prediction that the fluctuations
should derive from a Gaussian random-phase distribution.  All of our tests
have confirmed this prediction dramatically.

\section{Acknowledgments}
The WMAP datasets were developed by Princeton University and NASA's Goddard
Space Flight Center, and graciously provided to the public at
http://map.gsfc.nasa.gov/.  JRG thanks the National Science Foundation for
its support under grant AST 99-00772.

\vfill

\clearpage

\begin{figure}
%\epsscale{0.6}
%\plotone{colleyw_f1.eps}
%%\epsscale{0.9}
%%\plotone{full_stereo.pdf}
\caption{The WMAP data (internal linear combination) in the north (top) and
south (bottom) Galactic hemispheres, in stereographic projection, with the
Galactic Center at center.  The color map is a linear scale, chosen such
that hot spots are red and cold spots are blue with a total range in
temperature of $-200 \mu\mbox{K}$ to $+200 \mu\mbox{K}$.  The mean
temperature contour is in white.}
\label{fullstereo}
\end{figure}

\vfill

\clearpage

\begin{figure}
\epsscale{0.9}
\plotone{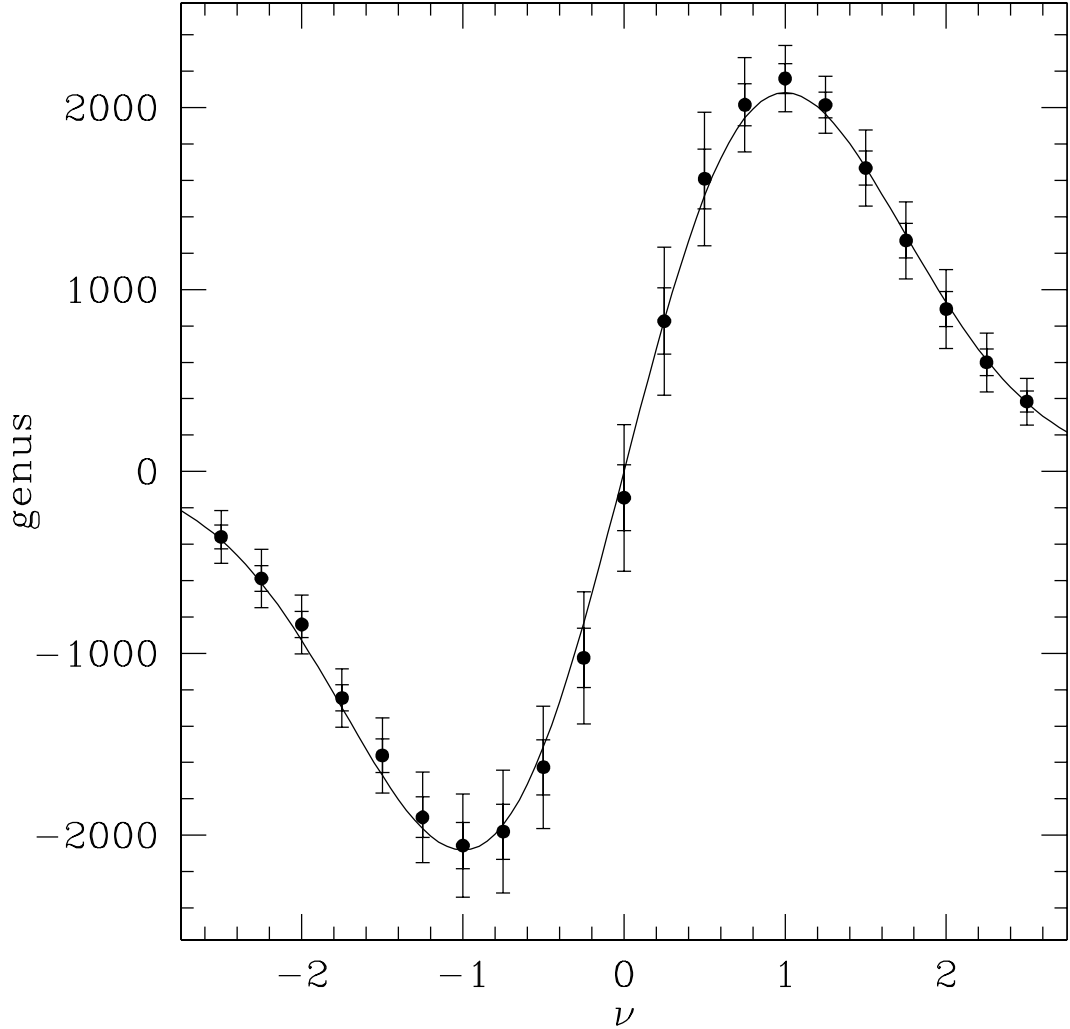}
%\plotone{hpgenus_2.pdf}
\caption{Total genus, $\gtwodeff$, for the original HEALPix projection of
the WMAP data (internal linear combination) for the entire sky, with no
masking.  The abscissa, $\nu$, is the departure from the pixel-wise
temperature mean for the whole sky in terms of pixel-wise temperature
standard deviation for the whole sky.  Inner errorbars are computed from
the standard deviation of the 12 HEALPix diamonds; the outer errorbars
reflect the sigma value scaled for 95.4\% confidence in the Student's-$t$
distribution.}
\label{hpg_2}
\end{figure}

\begin{figure}
\epsscale{0.9}
\plotone{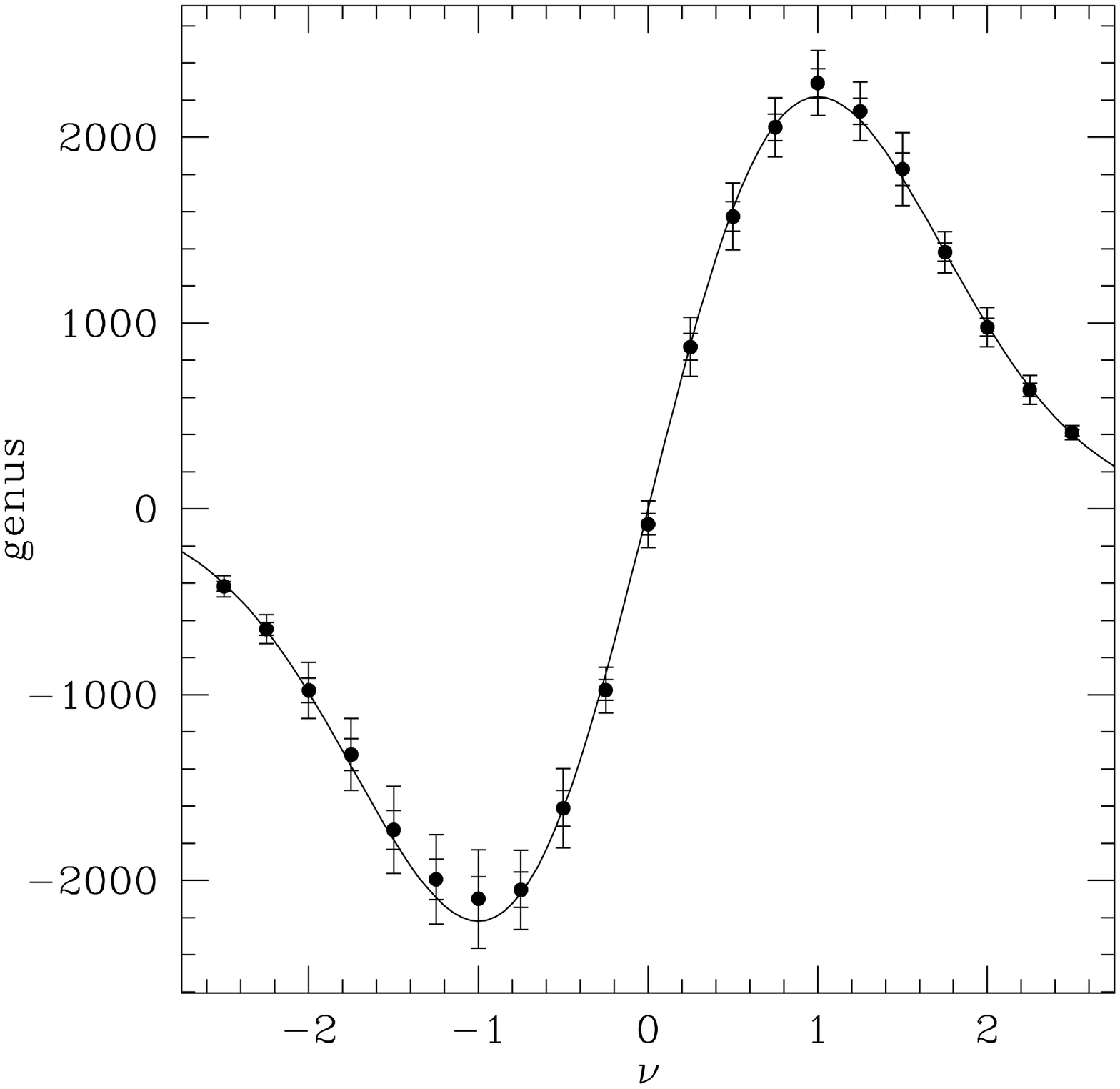}
%\plotone{hpgenus_1.pdf}
\caption{As in Fig.~\ref{hpg_2}, except the $\nu$ values are computed by
area fraction (see text).}
\label{hpg_1}
\end{figure}

\begin{figure}
\epsscale{0.9}
\plotone{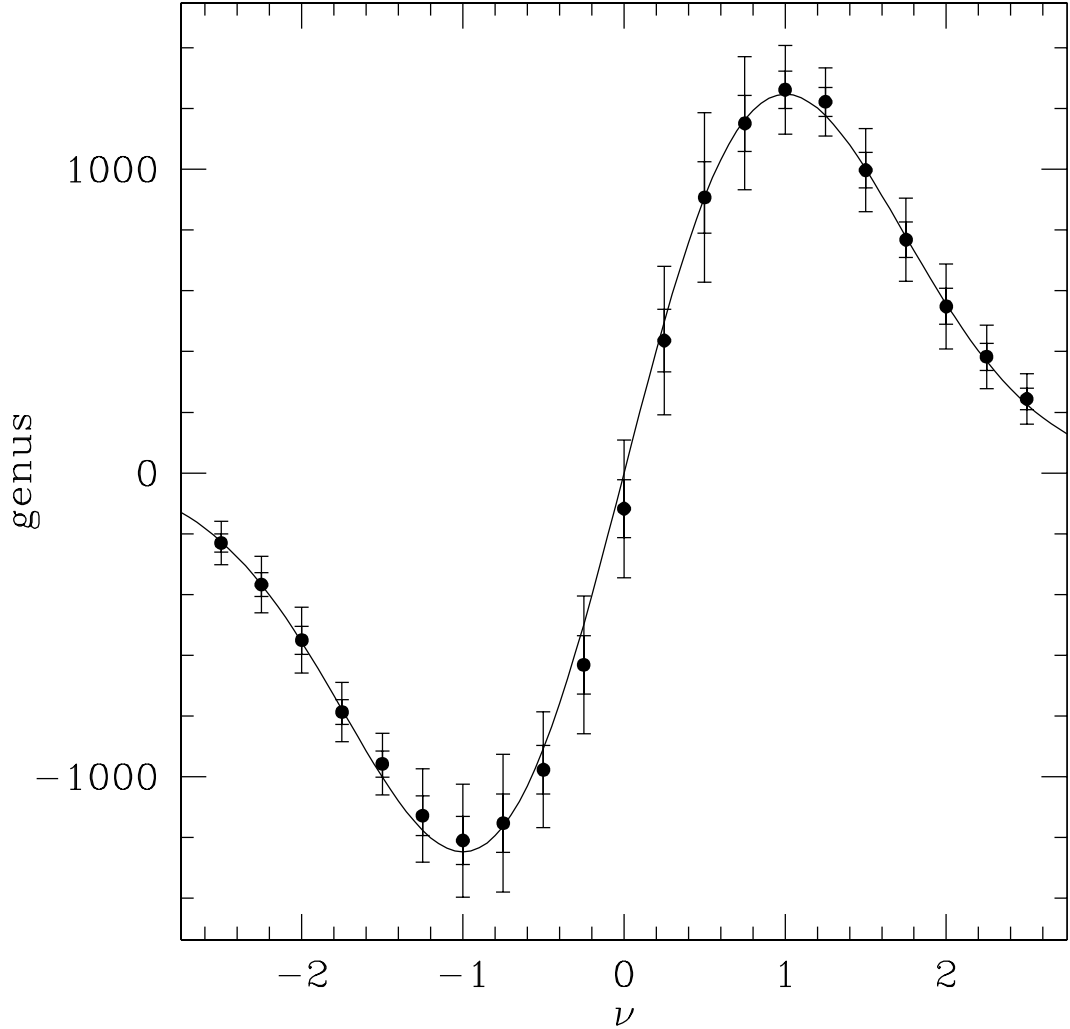}
%\plotone{genus_035_8_2.pdf}
\caption{Total genus, $\gtwodeff$, for the WMAP data (internal linear
combination), projected stereographically to a total smoothing of
$0.35^\circ$, then masked according to the {\it Kp0} standard. The
abscissa, $\nu$, is the departure from the pixel-wise temperature mean for
the whole sky in terms of pixel-wise temperature standard deviation from the
whole sky.  Inner errorbars are computed from the standard deviation of the
8 octants; the outer errorbars reflect the sigma value scaled for 95.4\%
confidence in the Student's-$t$ distribution.}
\label{g.35_2}
\end{figure}

\begin{figure}
\epsscale{0.9}
\plotone{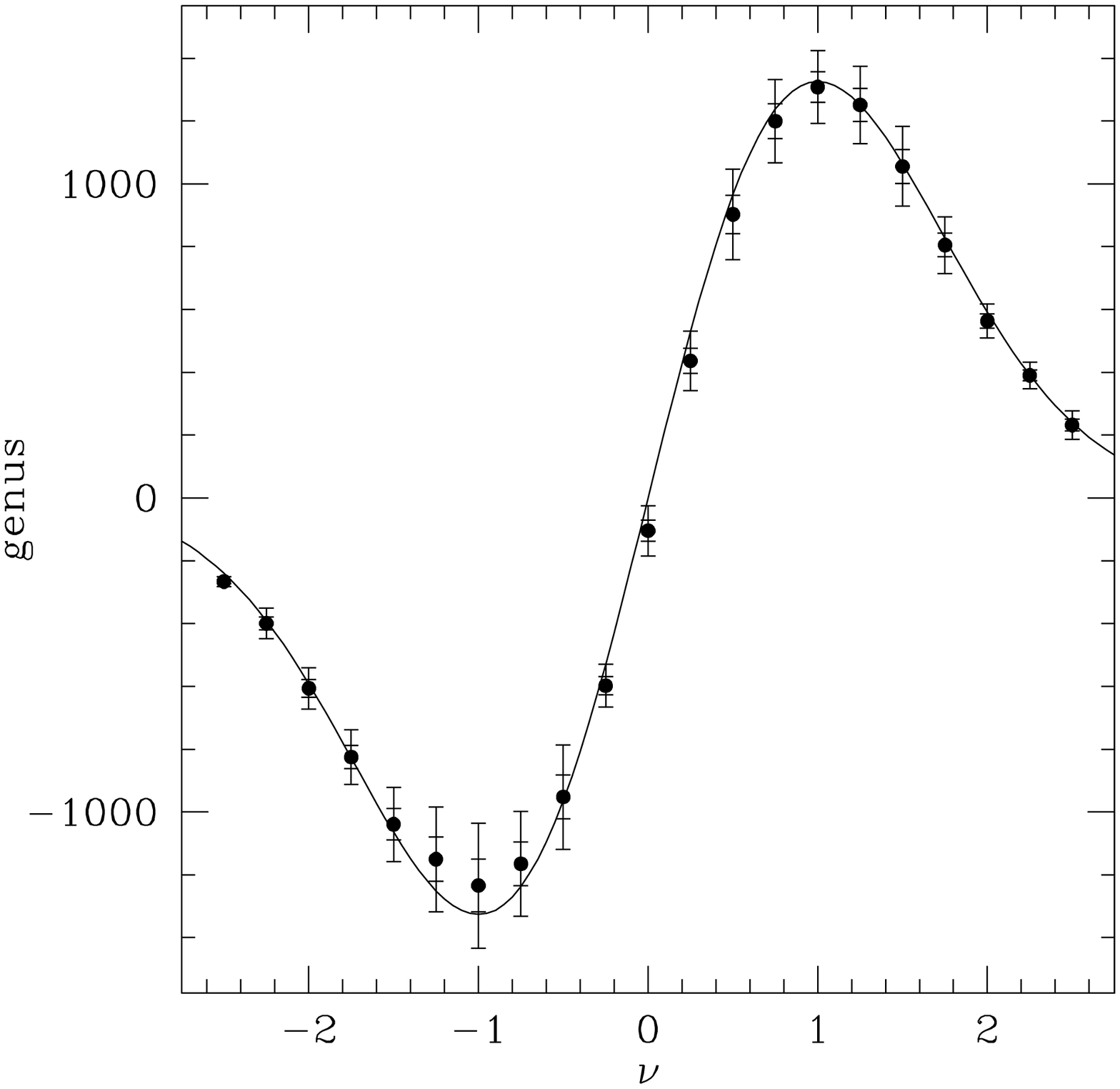}
%\plotone{genus_035_8_1.pdf}
\caption{As in Fig.~\ref{g.35_2}, except the $\nu$ values are computed by
area fraction (see text).}
\label{g.35_1}
\end{figure}

\begin{figure}
\epsscale{0.9}
\plotone{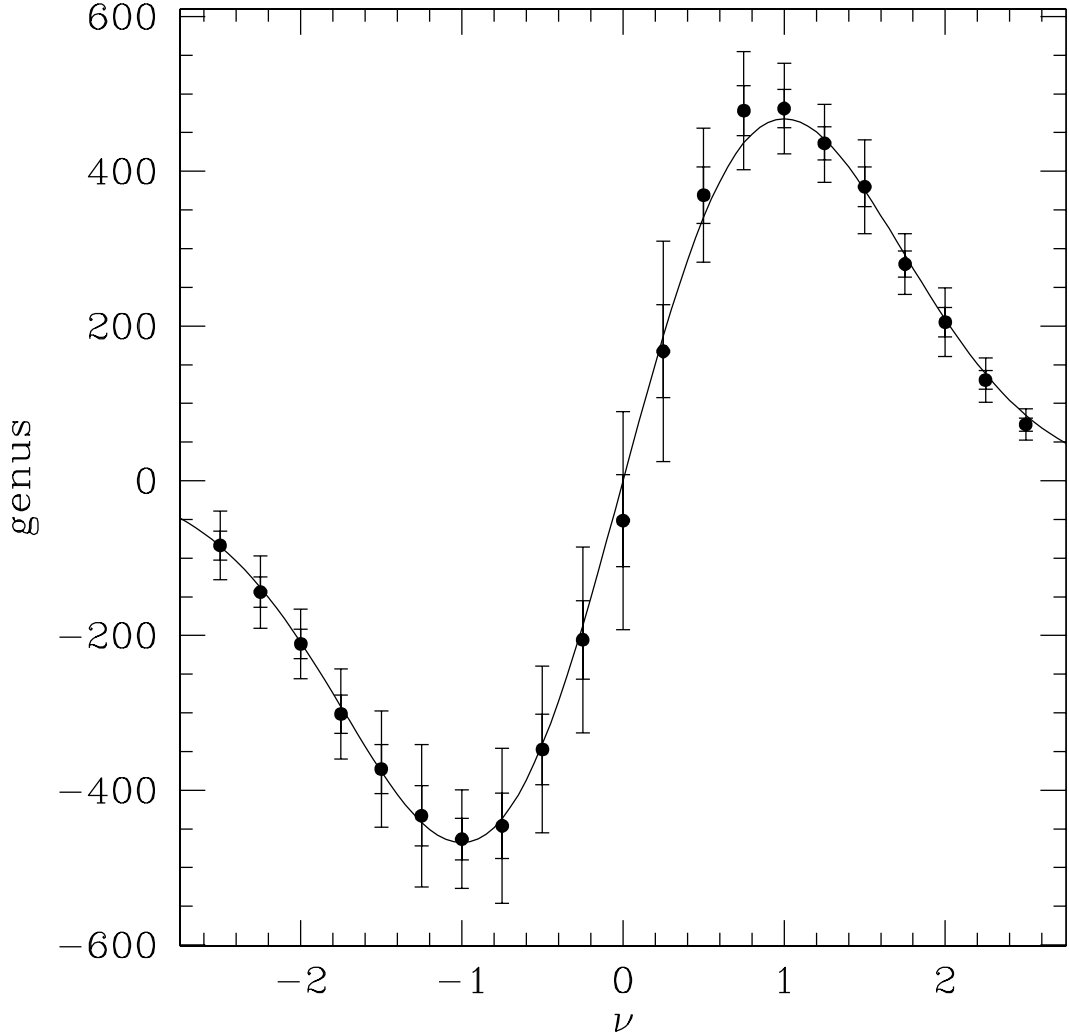}
%\plotone{genus_09_8_2.pdf}
\caption{Total genus, $\gtwodeff$, for the pre-smoothed WMAP data,
combined according to the internal linear combination coefficients, then
projected stereographically to a total smoothing of $0.9^\circ$, then
masked according to the {\it Kp0} standard.  The abscissa, $\nu$, is the
departure from the pixel-wise temperature mean for the whole sky in terms of
pixel-wise temperature standard deviation from the whole sky.  Inner
errorbars are computed from the standard deviation of the 8 octants; the
outer errorbars reflect the sigma value scaled for 95.4\% confidence in
the Student's-$t$ distribution.}
\label{g.9_2}
\end{figure}

\begin{figure}
\epsscale{0.9}
\plotone{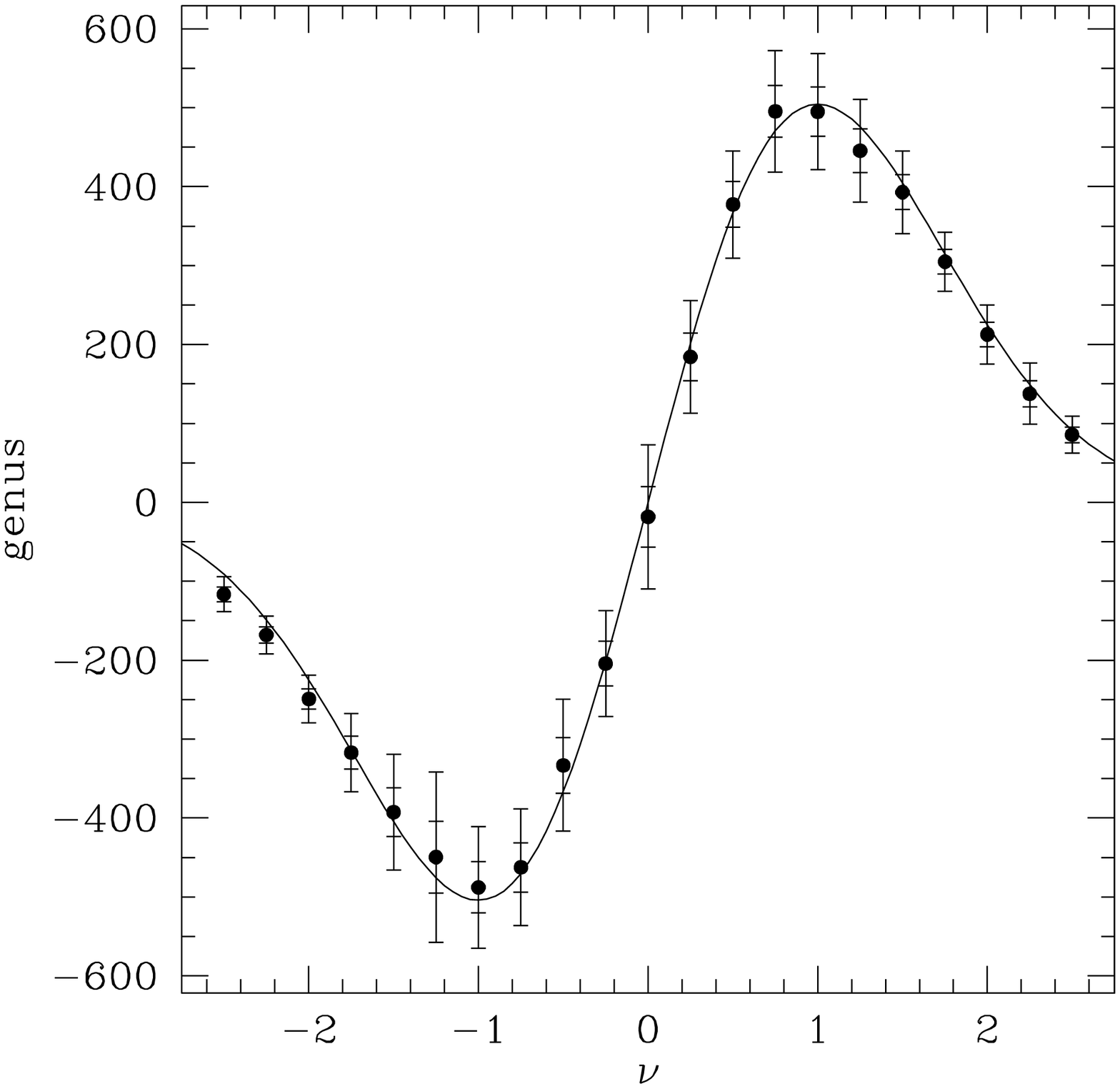}
%\plotone{genus_09_8_1.pdf}
\caption{As in Fig.~\ref{g.9_2}, except the $\nu$ values are computed by
area fraction (see text).}
\label{g.9_1}
\end{figure}

\end{document}